\documentclass[runningheads]{llncs}
\usepackage[T1]{fontenc}
\usepackage{graphicx}
\usepackage{soul}
\usepackage{color}
\usepackage{amsmath}
\usepackage{soul}
\usepackage{ulem}
\begin{document}
\title{HyPer-EP: Meta-Learning Hybrid Personalized Models for Cardiac Electrophysiology}
%
%
\author{Xiajun Jiang \inst{1} 
\and
Sumeet Vadhavkar \inst{1} 
\and
Yubo Ye \inst{2} 
\and
Maryam Toloubidokhti \inst{1} 
\and
Ryan Missel \inst{1} 
\and
Linwei Wang \inst{1} 
}
\authorrunning{Xiajun Jiang et al.}
%
\institute{Rochester Institute of Technology, Rochester, NY USA \and
Zheajiang University, Hangzhou, Zhejiang, China
}
%
\maketitle              
\begin{abstract}
Personalized virtual heart models have demonstrated increasing potential for clinical use, although the estimation of their parameters given patient-specific data remain a challenge. 
Traditional physics-based modeling approaches are computationally costly and often neglect the inherent structural errors in these models due to model simplifications and assumptions. 
Modern deep learning approaches, 
on the other hand, 
rely heavily on data supervision and lacks interpretability. 
In this paper, 
we present a novel hybrid modeling framework to describe a personalized cardiac digital twin as a combination of a physics-based known expression augmented by neural network modeling of its unknown gap to reality. We then present a novel meta-learning framework to enable the separate identification of both the physics-based and neural components in the hybrid model. 
We demonstrate the feasibility and generality of this hybrid modeling framework with two examples of instantiations and their proof-of-concept in synthetic experiments.
\end{abstract}
\section{Introduction}
Personalized virtual heart models, 
such as those describing the electrphyisolgical (EP) process of the heart, have shown significant progress in risk stratification \cite{arevalo2016arrhythmia}, treatment planning \cite{zahid2016feasibility}, and outcome predictions \cite{SERMESANT2012201}. Effective personalization of virtual heart models, 
especially estimation of model parameters pertaining to patient-specific tissue properties, 
however remain a critical challenge 
due to the ill-posed nature of the inverse problem, 
the myriad of modeling assumptions involved, 
and the computational cost associated with these models. 

Many efforts have been devoted to  personalize the parameters of virtual EP models of the heart. 
Earlier works have focused on iterative optimization/inference to 
minimize the discrepancy between model outputs and measured data \cite{sermesant2012patient,wong2015velocity}. 
Despite significant progress, 
the iterative nature of these approaches 
involving multiple runs of the EP model 
make it less appealing for clinical use. 
More importantly, 
they attribute discrepancies between model outputs and data observations solely to the model parameters being optimized, essentially assuming the absence of structural or other errors in the model.  
This overlooks unknown errors within a virtual EP model associated with its structural assumptions and simplifications, which may in turn lead to erroneous parameter identification if this unknown error is non-negligible. 
We refer to this as \textit{white-box} models. 

Recent machine learning (ML) and deep learning (DL) advances have brought successes in the personalization of virtual heart models. 
Examples include learning input-output relationship between the parameters and outputs of the EP model \cite{kashtanova2021ep}, 
or the recent meta-learning approach to learn to adapt a neural-network as a surrogate for the EP model \cite{10.1007/978-3-031-16452-1_5}. 
These data-driven approaches bypass the underlying physical principles governing cardiac EP process, 
but instead heavily rely on the availability of 
 large datasets  
on the input-output relationship being learned. 
Since these data (\textit{e.g.,} tissue property as model input, or spatiotemporal activity of action potential as model output) are not always
available in \textit{in-vivo} settings, 
most data-driven approaches resort to simulated data for supervision. As a result, 
its training involves computationally-prohibitive data generation, 
while its deployment to real data 
faces challenges of generalization.  
We refer to this as \textit{black-box} models which are further limited in interpretibility. 

To bridge the gap between white-box and black-box modeling, 
recent works have proposed the use of 
physics-informed neural networks (PINNs) in personalized virtual heart models \cite{10.3389/fcvm.2021.768419}. 
In PINNs, 
the neural network's output is governed by a partial differential equation (PDE) with known mathematical expressions, representing prior knowledge and removing the need of data supervison; 
the parameter of the PDE can be optimized at the same time as the PINN is being trained, 
achieving a personalized PINN and PDE at the same time. 
However, 
although \textit{informed} by a given PDE in the training loss, 
the PINN is still a black-box function; 
moreover, the constraining PDE  
assumes exact and perfect mathematical descriptions of the underlying system of interest: 
this \textit{parallel} integration of white- and black-box modeling thus unfortunately inherits their respective limitations: 
the PINN is 
with limited interpretibility, 
while its personalization may suffer if the constraining white-box models are imperfect. 
Additionally, 
the joint optimization of the PINN and PDE parameter have to be performed for each individual subject, 
limiting its clinical applicability. 


To overcome the above challenges, we propose a novel
hybrid modeling approach towards personalized virtual heart models, replacing existing 
\textit{white-box} or \textit{black-box} modeling approaches with a \textit{gray-box} approach. 
Unlike existing \textit{physics-informed} gray-box models, 
we further move towards a \textit{physics-integrated} gray-box model that 
explicitly hybridize physiological models with neural-network models inside the digital twin.
While the concept of hybrid models has emerged in various domains including virtual heart models,\cite{ALPS,NeuralSim,UDE,Inria},
a critical roadblock is the assumption of direct supervision on the variables being modeled, which is not applicable in virtual heart models where the modeled 
variable (\textit{e.g.,} spatiotemporal propagation of action potentials in the heart) is often only partially or indirectly observed. 
To this end, 
we further 
addresses the challenge of \textit{unsupervised} identification of hybrid models by a novel meta-learning strategy 
to separately identify the parameters of the physiological model and its gaps to observed data.
At training time, 
the proposed method 
for meta-learning hybrid personalized cardiac EP (HyPer-EP) models 
does not require ground truth of the variables being modeled, leveraging prior physiology while learning to identify its gap to observed data. 
At test time, 
HyPerEP enables 
personalization of a hybrid cardiac digital twin -- composed of an interpretable physiological component and a neural components accounting for its errors -- using 
subject-specific data via efficient feedforward computations. 
We demonstrate the feasibility and generality of HyPer-EP with two examples of instantiations, providing evidence for their 
feasibility and benefits over physics-based or neural modeling alone in 
synthetic experiments.

\section{Problem Formulation}
\label{sec:bck}



Consider the goal of obtaining a personalized model $\mathcal{M}(\mathbf{\theta})$ describing the EP process of the ventricles in the form of the spatiotemporal propagation of action potentials 
$\mathbf{x}_{0:T}$, 
with patient-specific parameter $\theta$ and observations $\mathbf{y}_{0:T} = g(\mathbf{x}_{0:T})$.


In a white-box approach, $\mathcal{M}$ is a known mathematical expression $\mathcal{M}_{\text{PHY}}$ and, when given measurements $\mathbf{y}_{obs}$, the value of $\theta$ is optimized 
to fit the output of $\mathcal{M}_{\text{PHY}}$ to $\mathbf{y}_{obs}$ via data-fitting metrics such as mean-squared-errors (MSE):
\begin{equation}
    \hat{\theta} = \arg\min_{\theta} ||  g(\mathcal{M}_{\textrm{PHY}}(\mathbf{\theta})) - \mathbf{y}_{obs} ||_2^2
\end{equation}
where all potential errors in $\mathcal{M}_{\text{PHY}}$ due to model assumption and simplifications are neglected, and its difference from $\mathbf{y}_{obs}$ is solely attributed to parameter $\theta$.


In a black-box approach, 
$\mathcal{M}$ is often a deep neural network (DNN) $\mathcal{M}_{\phi}$ and its weight parameters $\phi$ is typically learned given a large number of paired data $\{\theta^i, \mathbf{x}_{0:T}^i\}_{i=1}^N$ in a supervised loss using, for instance, MSE: 
\begin{equation}
    \hat{\phi} = \arg\min_{\phi} \sum_{i=1}^N|| \mathcal{M}_\phi(\theta^i) - \mathbf{x}_{0:T}^i ||_2^2
\end{equation}
where $\{\theta^i, \mathbf{x}_{0:T}^i\}_{i=1}^N$ are often obtained via simulation data because they are not readily available in practice, 
raising challenges of generalization to real data. 


In the recently-emerged PINN approach, a DNN $\mathcal{M}_\phi$ with weight parameters $\phi$, is supervised by a partial differential equation (PDE) $\mathcal{M}_{\text{PHY}}$ with known mathematical expressions and potentially unknown parameter $\theta$. Given available data on 
$\mathbf{x}_{0:T}$, both $\phi$ and $\theta$ of the two models can be simultaneously optimized:
\begin{equation}
    \{\hat{\phi}, \hat{\theta} \} = \arg\min_{\phi,\theta} \{ 
    ||\mathcal{M}_\phi - \mathbf{x}_{0:T} ||_2^2 + 
    \lambda ||\mathcal{M}_{\text{PHY}}
    (\mathcal{M}_\phi; \theta)||_2^2 \}
\end{equation}
where the first term fits $\mathcal{M}_\phi$'s output to available data
(data-fitting loss), and 
the second term encourages $\mathcal{M}_\phi$'s output to follow the governing PDE specified by $\mathcal{M}_{\text{PHY}}$ (PDE residual loss). 
With this \textit{parallel} integration, 
 $\mathcal{M}_\phi$ is still a blackbox while $\mathcal{M}_{\text{PHY}}$ is assumed to represent exact knowledge of the system. 
Furthermore, because $\theta$ is unique to each individual, this joint optimization of $\phi$ and $\theta$ must be repeated for each given set of observations $\mathbf{x}_{0:T}$.

\section{Methodology}

In this work, 
we propose a novel 
Hybrid Personalized (HyPer) modeling framework to  
address the limitations of purely white-box or black-box models  while marrying their respective strengths. 
Unlike the parallel integration in current \textit{physics-informed} approaches as described in Section \ref{sec:bck}, 
HyPer is underpinned by a \textit{physics-integrated} hybrid model $\mathcal{M}_{\text{Hybrid}}$ consisting of known mathematical expression 
$\mathcal{M}_{\text{PHY}}$ 
augmented by an unknown neural component $\mathcal{M}_\phi$ to account for its potential gap to reality, 
each parameterized by learnable parameters. 
This hybrid model is then situated within the latent space of an encoding-decoding architecture 
to bridge the variables being modeled to their indirect observations at the data space, 
enabling a novel \textit{unsupervised} learning paradigm with a learn-to-identify meta-learning formulation to address the identifiability issue associated with the separate identification of 
$\mathcal{M}_{\text{PHY}}$ 
and $\mathcal{M}_\phi$ in the hybrid model. 
This hybrid generative modeling 
and the learn-to-identify inference strategy constitute the backbone in HyPer, 
which we elaborate below in the context of cardiac EP models (and referred to as HyPer-EP).



\subsection{Hybrid Modeling of Cardiac EP Process}

\subsubsection{General HyPer formulation:}
The proposed hybrid model $\mathcal{M}_{\text{Hybrid}}$ is a combination of known mathematical expression $\mathcal{M}_{\text{PHY}}$ 
and an unknown neural function $\mathcal{M}_\phi$, where the latter is intended to capture potentially unmodeled complexities or errors inherent in the simplified white-box representation $\mathcal{M}_{\text{PHY}}$: 
\begin{equation}
\label{eqn:hybrid}
      \mathcal{M}_{\text{Hybrid}} = \mathcal{M}_{\text{PHY}} + \mathcal{M}_\phi     
\end{equation}
Note that Equation \eqref{eqn:hybrid} denotes a general framework where the hybridization of $\mathcal{M}_{\text{Hybrid}}$ and $\mathcal{M}_{\text{PHY}}$ can be realized in various strategies. Below we present two examples of instantiations of  $\mathcal{M}_{\text{Hybrid}}$ in the context of cardiac EP modeling.
\subsubsection{Instantiation 1 -- HyPer to bridge simple and data-generating physics:} 
In this instantiation, 
we consider the single-variable
Eikonal model as $\mathcal{M}_{\text{PHY}}$ due to its popularity associated with its simplicity 
and fast computation for personalized EP modeling. While 
the Eikonal model computes only the arrival time of activation wavefront in space, we use $\mathcal{M}_\phi$ to bridge its gap to the spatiotemporal action potential depolarization and repolarization process. 

\uline{$\mathcal{M}_{\text{PHY}}$:} 
We consider the isotropic but heterogeneous Eikonal PDE given by:
\begin{equation}
\label{eqn:phy}
    |\nabla T(\mathbf{r})|\theta(\mathbf{r}) = 1
\end{equation} 
where $T(\mathbf{r})$ denote the arrival time of activation wavefront at spatial location $\mathbf{r}$, and $\theta(\mathbf{r})$ denotes the local conduction velocity at $\mathbf{r}$. 
Given the initial locations of electrical activation and the heterogeneous conduction velocity $\theta(\mathbf{r})$ across the myocardium of the heart, 
Equation \eqref{eqn:phy} can be solved in real-time to describe the 
isotropic propagation of action potential wavefront through the myocardium. It however does not model the realistic depolarization and repolarization dynamics of local action potential, nor the anisotropic spatial diffusion due to fiber orientation, which will be accommodated in an unknown neural component.

\uline{$\mathcal{M}_\phi$:} 
We model $\mathcal{M}_\phi$ to take inputs from Eikonal's output $T(\mathbf{r})$ and convert it to action potential $\mathbf{x}_{0:T}$ throughout the myocardium over time $[0, T]$:
\begin{equation}
    \mathbf{x}_{0:T} = \mathcal{M}_\phi(T(\mathbf{r}))
\end{equation}

Because $\mathbf{x}_{0:T}$ lives on a 3D geometry of the heart, 
we represent the myocardial mesh with an undirected  k-nearest-neighbors (kNN) graph: 
each node of the myocardial mesh represents a vertex in the graph, and an edge is formed between a mesh node and its k nearest node neighbors as measured by Euclidean distance; the edge  
attributes between vertices are defined as the normalized differences in their 3D coordinates if an edge exists. 
On a given graph, 
$\mathcal{M}_\phi$ is realized as a spatial-temporal graph convolutional neural network (ST-GCNN) builded on the spline-GCNN \cite{GCNN} with interlaced graph convolution and temporal feature extraction operations: 
\begin{equation}
    (\mathbf{f} \ast \mathbf{g})=\sum_{j\in N(i)} \mathbf{f}(j)\cdot \sum_{\mathbf{p}\in \mathcal{P}} \omega_\mathbf{p}B_\mathbf{p}(\mathbf{u}(i,j))
\end{equation}
where $\mathbf{f}$ is graph node features at each time instant, $\mathbf{u}(i,j)$ is the edge attribute between vertex $i$ and $j$, $\mathbf{g}(\cdot)$ is the convolution kernel, $B_{\mathbf{p}}(\cdot)$ is the spline basis with its the Cartesian product $\mathcal{P}$ and $\omega_{\mathbf{p}}$ are trainable parameters. And temporal feature extraction operation is implemented by the fully connected layers.
The choice of spline-GCNN with its spatial convolution kernels allows the modeling across different hearts at both training and test time. 

This hybrid formulation leverages the fast conduction physics described by the Eikonal model, while allowing data-driven modeling of its gap to reality. Once $\theta(\mathbf{r})$ and $\phi$ are each identified for $\mathcal{M}_{\text{PHY}}$ and $\mathcal{M}_{\phi}$, respectively, 
a personalized hybrid cardiac EP model will be obtained.

\subsubsection{Instantiation 2 --  HyPer as a universal differential equation (UDE):}  
In this instatiation, 
we model the PDE of action potential $\frac{d\mathbf{x}_t}{dt}$ 
with a combination of a known mathematical expression $f_{\textrm{PHY}}$ and an unknown neural function $f_{\textrm{NN}}$:
\begin{equation}
\label{eqn:hybrid}
    \frac{d\mathbf{x}_t}{dt} 
    =f_{\textrm{PHY}}(\mathbf{x}_t; \theta)+f_{\textrm{NN}_\phi}(\mathbf{x}_t)
\end{equation}
where $f_{\textrm{PHY}}(\mathbf{x}_t; \theta)$ represents a known EP model with unknown parameter $\theta$, and  
$f_{\textrm{NN}_\phi}(\mathbf{x}_t)$ represents its potential errors. 

\uline{$f_{\textrm{PHY}}$:} In 
real-data experiments, 
we consider $f_{\textrm{PHY}}$ to be the two-variable \textit{Alieve-Panfilov} model describing spatiotemporal action-potential generation \cite{aliev1996simple}. 
In synthetic data experiments, as a proof-of-concept,  
we consider $f_{\textrm{PHY}}$ to be the \textit{Alieve-Panfilov} with a missing term to represent its structural error to the data-generating full \textit{Alieve-Panfilov} model. 

\uline{$f_{\textrm{NN}}$:} 
To demonstrate the HyPer-EP is a general framework that is agnostic to the type of physics-based or nueral-network functions used, here we model $f_{\textrm{NN}}$ as a MLP. The MLP is a fully connected neural network with a ReLU activation at each layer and a Tanh activation function at the final layer. The input to the network is the transmembrane potential 'u' and the recovery term 'v'. The MLP aims to correct the missing repolarisation due to partial physics. 

This hybrid formulation 
learns a hybrid PDE that is partly known in structure but with an unknown parameter $\theta$ that is globally homogeneous, and a partly unknown component represented by $f_{\textrm{NN}_\phi}$: when both  $\theta$ ad $\phi$ are identified, a hybrid cardiac EP model is obtained.







\subsection{Learning to Identify}

The identification of $\mathcal{M}_{\text{Hybrid}}$ requires simultaneous identification of the parameter of $\mathcal{M}_{\text{PHY}}$ and $\mathcal{M}_{\text{NN}}$. Formally, we cast this into a meta-learning formulation. Consider a dataset $\mathcal{D}$ of action potentials with $M$ similar but distinct underlying dynamics: $\mathcal{D}=\left \{ \mathcal{D}_j \right \}_{j=1}^{M}$. For each $\mathcal{D}_j$, we consider disjoint few-shot context instances $\mathcal{D}_j^s=\left \{ \mathbf{y}_{0:T}^{s,1}, \mathbf{y}_{0:T}^{s,2}, \dots,\mathbf{y}_{0:T}^{s,k} \right \}$ and query instances $\mathcal{D}_j^q=\left \{ \mathbf{y}_{0:T}^{q,1}, \mathbf{y}_{0:T}^{q,2}, \dots,\mathbf{y}_{0:T}^{q,d} \right \}$, where $k \ll d$. Then we formulate a meta-objective to learn to identify the underlying true parameter vector $\theta$ of $\mathcal{M}_{\text{PHY}}$ from \textit{k}-shot context instances $\mathcal{D}_j^s$, such that the identified HyPer is able to forecast for any query instances in $\mathcal{D}_j^q$ given only an estimate of its initial state $\hat{\mathbf{x}}_{0,j}^q$. More specifically, we have a feedforward meta-model $\mathcal{G}_\zeta(\mathcal{D}_j^{s})$ to learn to identify $\theta$ for dynamics $j$ as:
\begin{equation}
    \hat{\theta}_j=\mathcal{G}_\zeta(\mathcal{D}_j^{s})=\frac{1}{k}\sum_{\mathbf{x}_{0:T}^s\in \mathcal{D}_j^s}\mathcal{\xi}_\zeta(\mathbf{y}_{0:T}^s)
\end{equation}
where an embedding is extracted from each individual context instance via a 
meta-encoder $\mathcal{\xi}_\zeta$ and gets aggregated across $\mathcal{D}_j^{s}$ to extract knowledge shared by the set. $k$ is the size of the context set, and its value can be fixed or variable which we will demonstrate in the ablation study.

Given the inferred $\hat{\mathbf{x}}_{0,j}^q$ and $\theta$, we minimize the forecasting accuracy on the query instances. 
\begin{equation}
    \left \{ \hat{\phi},\hat{\zeta} \right \}=\arg \min_{\phi,\zeta} \sum_{j=1}^M\sum_{\mathbf{y}_{0:T}^q\in\mathcal{D}_j^q}\left \| \mathbf{y}_{0:T}^q - g(\hat{\mathbf{x}}_{0:T}^q) \right \|_2^2 
\end{equation}

\section{Experiments and Results}


Our proof-of-concept experiments were run on synthetic data generated by the 
two-Variable Aliev-Panfilov model for both instantiations.  

\begin{equation}
\label{full}
\begin{aligned}
 \frac{du}{dt} &= \nabla (D\nabla u) + k*u(1-u)*(u-a) - uv\\
  \frac{dv}{dt} &= -e(k*u(u-a-1)+v), 
\end{aligned}
\end{equation}
where $u$ represents action potential, $v$ recovery current, 
$D$ the conductivity tensor, 
and the rest of the parameters controlling the temporal shape of the action potential. In particular, 
parameter $a$ is known to control the excitability of heart tissue, where an increased value of $a$ results in a reduced action potential duration and amplitude until an inability to activate. 
For diversity, 
in instatiation 1, we consider parameter $a$ as spatially-varying to mimic regions of infarcted tissue for different subjects; 
in instatiation 2, we consider parameter $a$ as spatially-homogeneous but with different values for different subjects.

 \begin{figure*}[t]
     \centering
     \includegraphics[width=.9\linewidth]{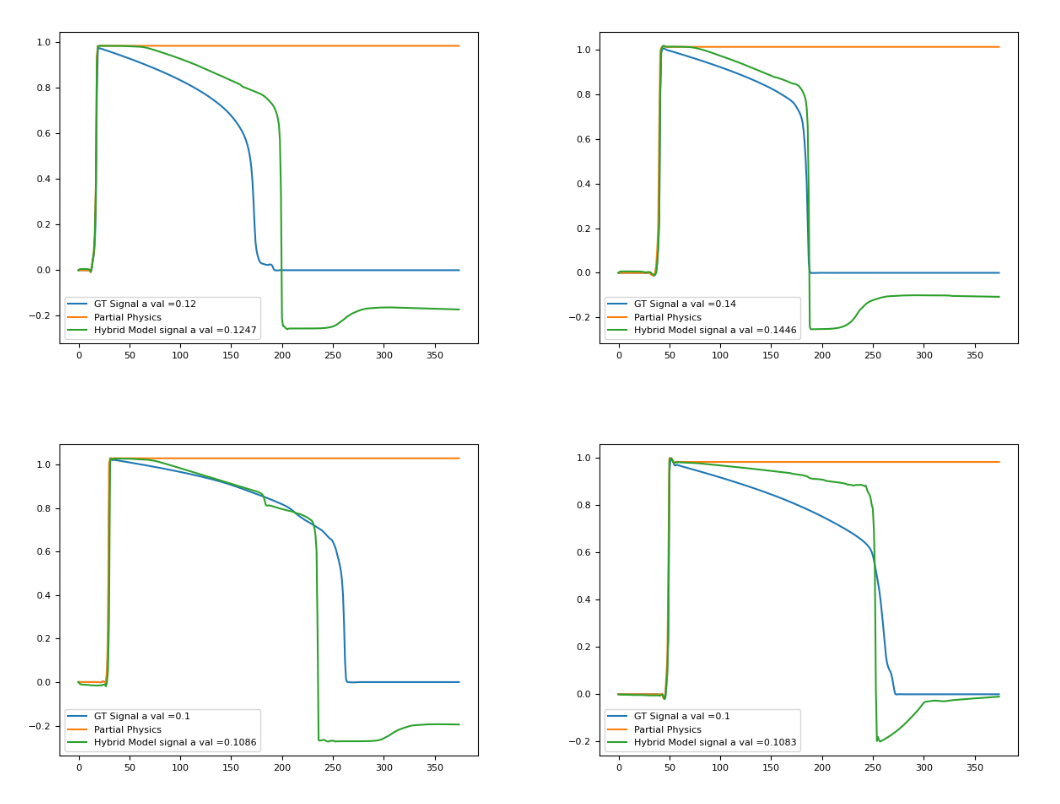}
     \caption{Visual examples of the action potential obtained from $\mathcal{M}_{\text{PHY}}$ versus $\mathcal{M}_{\text{Hybrid}}$ in comparison to the truth, along with the true and estimated values for parameter $a$.}
     \label{fig:result1}
 \end{figure*}

\subsubsection{Results on Instantiation 1:} 
We use the signal generated as by the full Aliev-Panfilov model (Equation \eqref{full}) as the ground truth of the experiments.
The signal is generated on 1862 volumetric heart meshes with 186different points of activation, 
each repeated for how many different parameter settings. 

In HyPer-EP, as described earlier the partial physics $\mathcal{M}_{\text{PHY}}$ represents the Eikonal model and   
$\mathcal{M}_{\text{NN}}$ that 
has two linear layers followed by three layers of interlaced graph convolution for spatial features and 1D convolution to recover the signal from the activation time map. 
The meta-encoder is modeled using
another three layers of interlaced graph convolution to extract spatial features and 1D convolution to aggregate temporal features,
then averaging on all context samples to estimate the parameter mask. 
Hyper-EP is trained to, 
given the initial excitation point of a query sample 
and parameter $\theta$ estimated from 
$k=5$ context samples, 
to reconstruct the action potential sequence for the query example.
Hyper-EP was trained on three parameter settings in total with around 200 data samples, and tested on five parameter settings with around 60 data samples in each parameter setting.

\begin{figure*}[t]
    \centering
    \includegraphics[width=.9\linewidth]{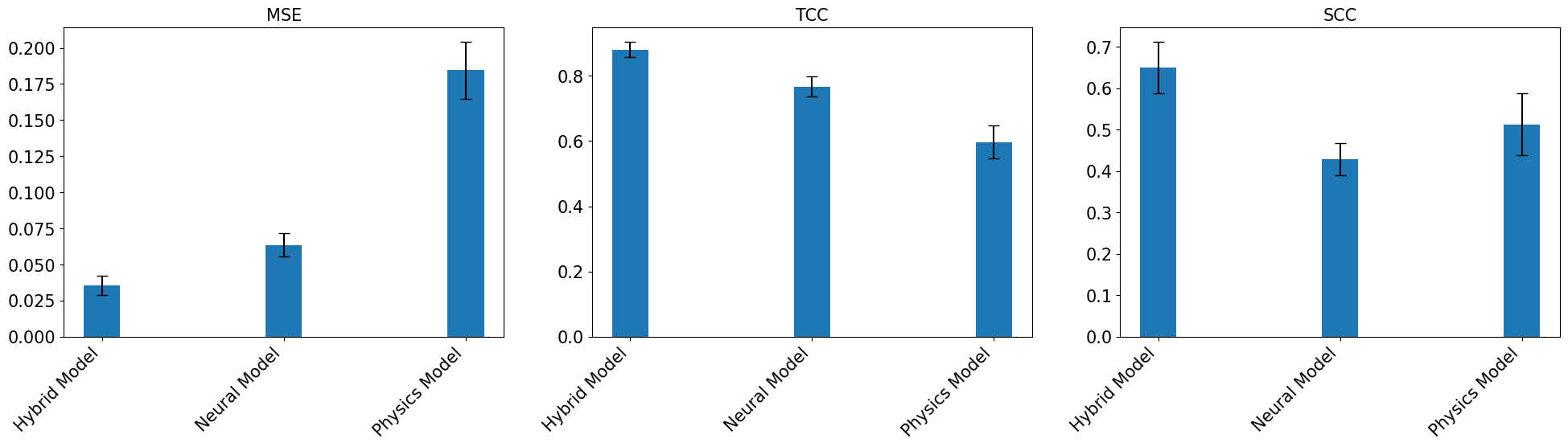}
    \caption{Summary of signal reconstruction performance on $\mathcal{M}_{\text{Hybrid}}$, $\mathcal{M}_{\text{NN}}$, and $\mathcal{M}_{\text{PHY}}$.}
    \label{fig:ins1_metric}
\end{figure*}

\begin{figure*}[t]
     \centering
    \includegraphics[width=.9\linewidth]{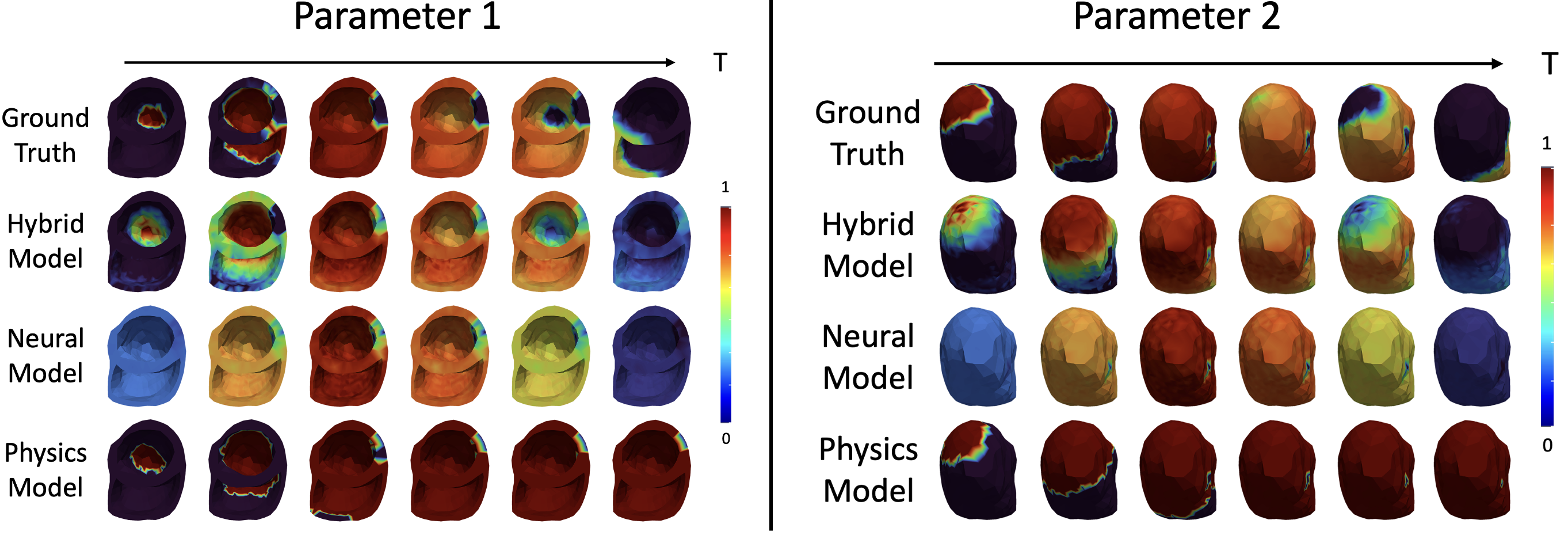}
     \caption{Visual examples of the action potential obtained from $\mathcal{M}_{\text{Hybrid}}$ versus $\mathcal{M}_{\text{NN}}$ and $\mathcal{M}_{\text{PHY}}$.}
     \label{fig:ins1_visual}
 \end{figure*}

Fig.~\ref{fig:ins1_metric} compares the performance of using only $\mathcal{M}_{\text{PHY}}$, 
$\mathcal{M}_{\text{NN}}$, 
and $\mathcal{M}_{\text{Hybrid}}$ in the same meta-learning framework, 
considering metrics of mean squared error (MSE), 
spatial correlation coefficient (SCC), 
and temporal correlation coefficient (TCC) between the reconstructed and true action potential. 
Visual examples are shown in Fig.~\ref{fig:ins1_visual}. 
These results demonstrate the advantages of the hybrid model over either physics-based or neural-network modeling alone in learning personalized cardiac EP models.

\subsubsection{Results on Instantiation 2:} 
We use the signal generated as by the full Aliev-Panfilov model (Equation \eqref{full}) as the ground truth of the experiments.
The signal is generated on a volumetric heart mesh with 1862 different points of activation, 
each repeated for 4 different values of parameter $a$ (0.08, 0.10, 0.12 and, 0.14). 

In HyPer-EP, for the partial physics $\mathcal{M}_{\text{PHY}}$, we remove the $u*v$ term from Equation \eqref{full}) to be modeled by  
$\mathcal{M}_{\text{NN}}$ that 
has 2 linear layers with a sigmoid activation. 
The meta-encoder is modeled using 2 layers of LSTM cells with ReLU activations. 
Hyper-EP is trained to, 
given the initial excitation point of a query sample 
and parameter $a$ estimated from 
$k=10$ context samples, 
to reconstruct the action potential sequence for the query example.
Hyper-EP was trained 
with 1408 training samples and tested on 352 unique samples. 





HyPer EP is able to deliver a mean squared error (MSE) of $0.65*e ^{-5}$ for the identification of the excitability parameter for $\mathcal{M}_{\text{PHY}}$. 
While this identified partial physics $\mathcal{M}_{\text{PHY}}$ is only able to reconstruct the action potential with a large MSE and standard deviation (STD) of $0.38$, and, $0.42$ respectively   
$\mathcal{M}_{\text{Hybrid}}$ with the integration of $\mathcal{M}_{\text{NN}}$ is able to achieve a
reconstruction accuracy of $0.042$ with a STD of $0.19$. 
Fig.~\ref{fig:result1} provides visual examples of the action potential obtained from $\mathcal{M}_{\text{PHY}}$ versus $\mathcal{M}_{\text{Hybrid}}$ in comparison to the truth, 
providing evidence for HyPer-EP to adequately identify both the partially known physics and its gap to data. 

\section{Conclusion}

In this paper, 
we present a HyPer-EP framework that 
hybridizes physics-based modeling of prior knowledge with data-driven modeling of errors in prior physics, 
and demonstrates the feasibility of meta-learning how to identify both components to realize personalized hybrid modeling of cardiac EP. 
A proof of concept is presented on two examples of instantiations of HyPer-EP on synthetic data. 
Future works will investigate more extensive experimental evaluations as well as real-data use.

\bibliography{miccai2024}
\bibliographystyle{splncs04}

\end{document}